\documentclass[twocolumn,superscriptaddress,prd,showkeys,showpacs,nofootinbib]{revtex4-1}

\usepackage{graphicx}
\usepackage[colorlinks = true,
            linkcolor = cyan,
            urlcolor  = blue,
            citecolor = red,
            anchorcolor = blue]{hyperref}\usepackage{color}
\usepackage{amssymb}
\usepackage[nointegrals]{wasysym}
\usepackage{amsthm}
\usepackage{textcomp}
\usepackage{mathtools}

\usepackage[normalem]{ulem}

\usepackage{lineno}

\usepackage{comment}
\usepackage[separate-uncertainty,retain-explicit-plus,per-mode = symbol]{siunitx}
\usepackage{url}

\usepackage{float}
\newcommand{\0}{$0\nu\beta\beta$}

\begin{document}

\title{Report from the \emph{Workshop on Xenon Detector \0 Searches: Steps Towards the Kilotonne Scale}}
\author{A.~Anker}
\affiliation{SLAC National Accelerator Laboratory, Menlo Park, CA 94025, USA}
\author{A.~Avasthi}
\affiliation{Case Western Reserve University, Cleveland, OH 44106, USA}
 \author{M. Brodeur}
\affiliation{Department of Physics and Astronomy, University of Notre Dame, IN 46556, USA}
\author{T.~Brunner}
\affiliation{McGill University, Department of Physics, Montreal, QC, H3A 2T8, Canada}
\author{N.K.~Byrnes}
\affiliation{Physics Department, University of Texas at Arlington, Arlington, TX 76019, USA}
\author{N.R.~Catarineu}
\affiliation{Lawrence Livermore National Laboratory, Livermore, CA 944550, USA}
\author{A.~Cottle}
\affiliation{Department of Physics and Astronomy, University College London, London, WC1E 6BT, UK}
\author{P.~Englezos}
\affiliation{Rutgers University, Piscataway, NJ, 08854, USA}
\author{W.~Fairbank}
\affiliation{Physics Department, Colorado State University, Fort Collins, CO 80523, USA}
\author{D.~González Díaz}
\affiliation{Instituto Galego de Física de Altas Enerxías, University of Santiago de Compostela, Santiago de Compostela, 15782, Spain}
\author{R.~Guenette}
\affiliation{University of Manchester, Manchester, M13 9PL, UK}
\author{S.J.~Haselschwardt}
\affiliation{Lawrence Berkeley National Laboratory, Berkeley, CA 94720, USA}
\author{S.~Hedges}
\affiliation{Lawrence Livermore National Laboratory, Livermore, CA 944550, USA}
\author{M.~Heffner}
\email{heffner2@llnl.gov}
\affiliation{Lawrence Livermore National Laboratory, Livermore, CA 944550, USA}
\author{J.D.~Holt}
\affiliation{TRIUMF, Vancouver, BC V6T 2A3, Canada}
\author{A.~Jamil}
\affiliation{Physics Department, Princeton University, Princeton, NJ 08544, USA}
\author{B.J.P.~Jones}
\affiliation{Physics Department, University of Texas at Arlington, Arlington, TX 76019, USA}
\author{N.~Kawada}
\affiliation{Research Center for Neutrino Science, Tohoku University, Sendai 980-8578, Japan}
\author{S.~Leardini}
\affiliation{Instituto Galego de Física de Altas Enerxías, University of Santiago de Compostela, Santiago de Compostela, 15782, Spain}
\author{B.G.~Lenardo}
\email{blenardo@slac.stanford.edu}
\affiliation{SLAC National Accelerator Laboratory, Menlo Park, CA 94025, USA}

\author{A.~Marc}
\affiliation{Air Liquide Maritime SAS, 77550 Moissy-Cramayel, France}
\author{J.~Masbou}
\affiliation{Subatech, Université de Nantes, Nantes 44307, France}
\author{K.~Mistry}
\email{krishan.mistry@uta.edu}
\affiliation{Physics Department, University of Texas at Arlington, Arlington, TX 76019, USA}
\author{B.~Mong}
\affiliation{SLAC National Accelerator Laboratory, Menlo Park, CA 94025, USA}
\author{B.~Monreal}
\affiliation{Case Western Reserve University, Cleveland, OH 44106, USA}
\author{D.C.~Moore}
\affiliation{Wright Laboratory, Department of Physics, Yale University, New Haven, CT 06520, USA}
\author{D.R.~Nygren}
\affiliation{Physics Department, University of Texas at Arlington, Arlington, TX 76019, USA}
\author{I.~Olcina}
\affiliation{University of California, Berkeley, Department of Physics, Berkeley, CA 94720-7300, USA}
\affiliation{Lawrence Berkeley National Laboratory, Berkeley, CA 94720, USA}
\author{J.L.~Orrell}
\affiliation{Pacific Northwest National Laboratory, Richland, Washington 99352, USA}
\author{A.~Pocar}
\affiliation{University of Massachusetts, Amherst, Physics Department, Amherst, MA, 01003, USA}
\author{G.~Richardson}
\affiliation{Wright Laboratory, Department of Physics, Yale University, New Haven, CT 06520, USA}
\author{L.~Rogers}
\email{rogersl@anl.gov}
\affiliation{Argonne National Laboratory, Argonne, IL 60439, USA}
\author{R.~Saldanha}
\affiliation{Pacific Northwest National Laboratory, Richland, Washington 99352, USA}
\author{S.~Sangiorgio}
\affiliation{Lawrence Livermore National Laboratory, Livermore, CA 944550, USA}
\author{C.~Wittweg}
\affiliation{Physik-Institut, University of Zürich, 8057 Zürich, Switzerland}
\author{Q.~Xia}
\affiliation{Lawrence Berkeley National Laboratory, Berkeley, CA 94720, USA}
\author{L.~Yang}
\affiliation{Physics Department, University of California San Diego, La Jolla, CA 92093, USA}
\author{J.~Zennamo}
\affiliation{Fermi National Accelerator Laboratory (FNAL), Batavia, IL 60510, USA}
%

\date{\today}

\begin{abstract}
  These proceedings summarize the program and discussions of the  ``\textit{Workshop on Xenon Detector $0\nu\beta\beta$ Searches: Steps Towards the Kilotonne Scale}'' held on October 25-27 2023 at SLAC National Accelerator Laboratory. This workshop brought together experts from the communities of neutrinoless double-beta decay and dark matter detection, to discuss paths forward for the realization of monolithic experiments with xenon approaching the kilotonne scale.
\end{abstract}
\keywords{}

\maketitle



\section{Introduction}

The observation of neutrinoless double beta decay (\0), which would prove the neutrino is a Majorana fermion, is one of the most promising avenues for discovering beyond-the-standard-model (BSM) physics. It would be the first time the creation of matter would be observed in the laboratory by violating the long-observed lepton number conservation, would demonstrate a new mechanism of mass generation, and could provide two of the three Sakharov conditions (CP violation and baryon or lepton number violation) required for explaining the asymmetry between matter and antimatter in the Universe. In a general sense, \0 searches are the most powerful experimental test of lepton number violation (LNV), with sensitivities to new physics at energy scales beyond the reach of modern colliders~\cite{Dekens}. The sensitivity of \0 searches is typically quantified by assuming that the decay is mediated by the exchange of light Majorana neutrinos within the nucleus. In an effective field theory context, this corresponds to the sole dimension-5 LNV operator, the ``Weinberg'' operator, where the BSM physics is encoded in an ``effective'' neutrino mass $m_{\beta\beta}$ related to experiment through the equation: 
\begin{equation}
    T_{1/2}^{-1} = \mathcal{M}_{0\nu}^2 \, G_{0\nu} \, \frac{\langle m_{\beta\beta} \rangle^2}{m_e^2}.
\end{equation}
Here, $T_{1/2}$ is the \0 half-life, $\mathcal{M}_{0\nu}$ is the nuclear matrix element, $G_{0\nu}$ is a phase-space factor, and $m_e$ is the electron mass. While the theoretical uncertainty in this relationship is dominated by $\mathcal{M}_{0\nu}$, new developments in \emph{ab-initio} many-body methods promise to quantify and reduce this uncertainty in the near future~\cite{Holt}. The most sensitive searches for \0 to date establish a lower limit on the half-life of $T_{1/2} > \mathcal{O}(10^{26})$ years, corresponding to $m_{\beta\beta} \lesssim 100$~meV. 

The search for this ultra-rare process is being performed across multiple isotopes in several different detector types and will require a) near-complete background removal and b) enormous exposures of the relevant isotopes. 
The current and next generation of experiments are expected to achieve sensitivities to half-lives beyond $10^{28}$~yrs -- corresponding to just $\mathcal{O}(1)$ decay per year in $\mathcal{O}(1)$~tonne of isotope -- reaching a sensitivity to the effective neutrino mass around $\sim$10~meV. 
These experiments will probe deep into new parameter space, aiming to completely explore the $m_{\beta\beta}$ region in the inverted hierarchy scenario and testing a significant amount of new space in the normal hierarchy scenario under the minimal mechanism. 
However, to explore the remaining experimentally-accessible $m_{\beta\beta}$ parameter space, experiments will need to push down to $m_{\beta\beta} \approx 1$~meV, corresponding to half-lives of $\mathcal{O}(10^{30})$~years. 
This requires the observation of $\sim$1 decay per year in several hundred tonnes of isotopic mass, which would require experimental efforts to scale up by two orders of magnitude while also reducing backgrounds compared to present technologies.

\begin{figure*}[ht]
    \centering
    \includegraphics[width=0.9\textwidth]{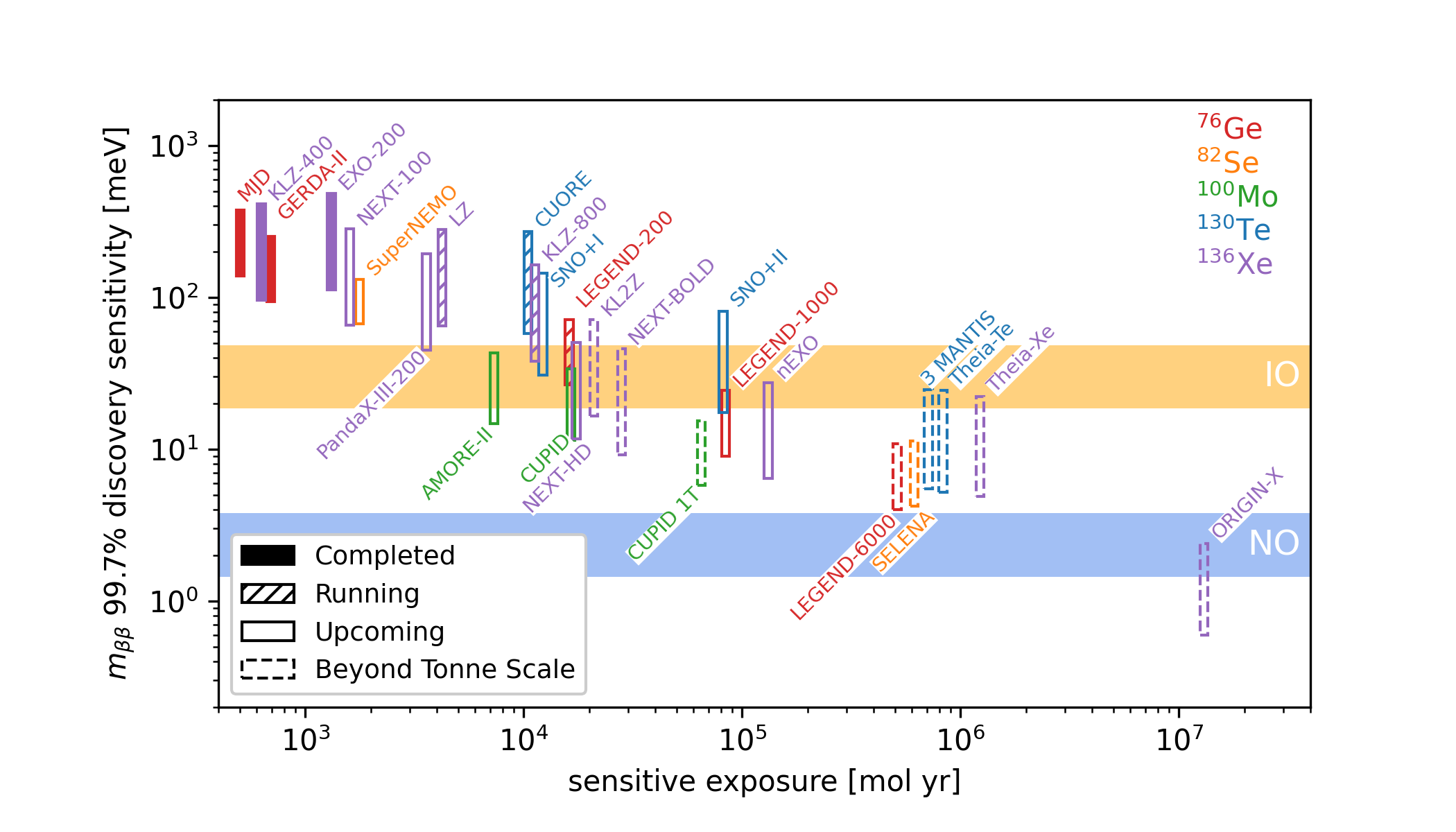}
    \caption{Sensitivity to the neutrino Majorana mass for various neutrinoless double beta decay detectors. The orange and blue bands correspond to the inverted and normal orderings respectively. Only experiments that can be built at the kilotonne scale will have the isotopic mass required to be sensitive to the meV mass scale~\cite{Heffner}.}
    \label{fig:originx}
\end{figure*}

Xenon-based experiments offer compelling paths toward reaching this goal. The \0 isotope $^{136}$Xe, which comprises 8.9\% of naturally occurring xenon, has a relatively high $Q$-value which boosts the decay rate by increasing the available phase space. Being gaseous at room temperature and chemically inert, xenon is straightforward to isotopically enrich. Most importantly, it can be deployed in multiple different detector technologies at large scales: the most sensitive (and largest-mass) search for \0 to date ($T_{1/2} > 2 \times 10^{26}$~yrs @ 90\% confidence level) comes from the KamLAND-Zen experiment using a $^{136}$Xe-doped liquid scintillator~\cite{Kawada}, and noble liquid and gas time projection chambers (TPCs) are under development for next- and beyond-the-next-generation experimental \0 programs. Recent work has shown that TPCs in particular -- either liquid or gas -- may indeed offer a technologically feasible path towards the 100-1000 tonne detector scale~\cite{PhysRevD.104.112007}. A consortium, ORIGIN-X, has been formed to foster collaboration among researchers across different projects with the common goal of realizing this ultimate \0 experiment~\cite{Heffner}. The projected sensitivity of such an experiment, compared to other existing and proposed programs, is illustrated in Figure~\ref{fig:originx}.  

Beyond \0, a carefully optimized kilotonne-scale xenon-based detector could also offer a rich scientific program in particle physics and astrophysics. Dual-phase xenon TPCs, for instance, are a leading technology in the search for WIMP dark matter, with planning underway for a Generation-3 (G3) experiment that would deploy a target mass of 40-80 tonnes of natural xenon~\cite{Kamaha}. A future kilotonne-scale detector, while background-limited by solar and atmospheric neutrinos' CE$\nu$NS interactions, could potentially achieve even greater sensitivities, or -- in the event of a G3 discovery -- provide precision measurements of the dark matter interaction. Furthermore, other isotopes of xenon are of interest for the study of second-order weak processes, and may be candidates for complementary LNV searches with detectors at kilotonne scales~\cite{Wittweg}. Finally, xenon detectors may offer unique insights into astrophysical neutrinos, with potential sensitivity to three separate interaction channels: CE$\nu$NS, electron elastic scattering, and charged-current neutrino-nucleus interactions~\cite{Haselschwardt}. A kilotonne scale detector could provide unprecedented spectral measurements of low-energy solar neutrinos~\cite{Richardson} and could play an important role in characterizing a supernova via its neutrino flux if one were to occur within a few kiloparsecs during its operation~\cite{Ghosh,Hedges}. 

Of course, there are substantial technical challenges in both isotopic acquisition and the development of a detector achieving near-perfect background rejection at such a large scale. The \emph{Workshop on Xenon Detector \0 Searches} brought together scientists from across the community to discuss beyond-tonne-scale detectors using xenon, with the goal of identifying key R\&D programs to open up the exciting scientific opportunities offered by a kilotonne-scale experiment.


\section{Xenon acquisition: status and challenges}

An immediate obstacle in progress towards kilotonne scale detectors is the availability of xenon as a raw material. Xenon is found primarily in the atmosphere, where it has a natural abundance of 87 parts-per-billion. Industrial use of xenon is dominated by semiconductor manufacturing, where XeF$_2$ is used for dry etching, with substantial markets also in spacecraft propulsion, lighting, medical applications, and others. Commercial supply of xenon gas is currently estimated at 65-100~\cite{Marc,Georgitzikis2022} tonnes per year and grows at an average rate of approximately 2 tonnes per year~\cite{Lindner}.

Xenon is currently produced as a byproduct of the production of liquid oxygen using Air Separation Units (ASUs), which perform cryogenic distillation. The process of separating oxygen from the air enriches the concentration of xenon in the output stream by several orders of magnitude, effectively subsidizing the cost of the subsequent production of pure xenon by $\sim$90\%~\cite{Sangiorgio}. This results in a highly inelastic supply. To extract and purify the xenon from the output stream, specialized process equipment must be added to the ASU. Only a fraction of the $\sim$500 ASUs worldwide have this capability, as there is a significant up-front capital cost $\mathcal{O}$(\$100~M) for installing the necessary infrastructure. Industrial xenon consumers typically guarantee fixed prices by establishing long-term contracts that, if necessary, include the cost of building up production capacity to guarantee that the product can be delivered. The result is that, of the 65-100t/yr produced worldwide, only $\sim$10\% is typically available on an open market~\cite{Lindner}. 

\begin{figure}
    \centering
    \includegraphics[width=0.49\textwidth]{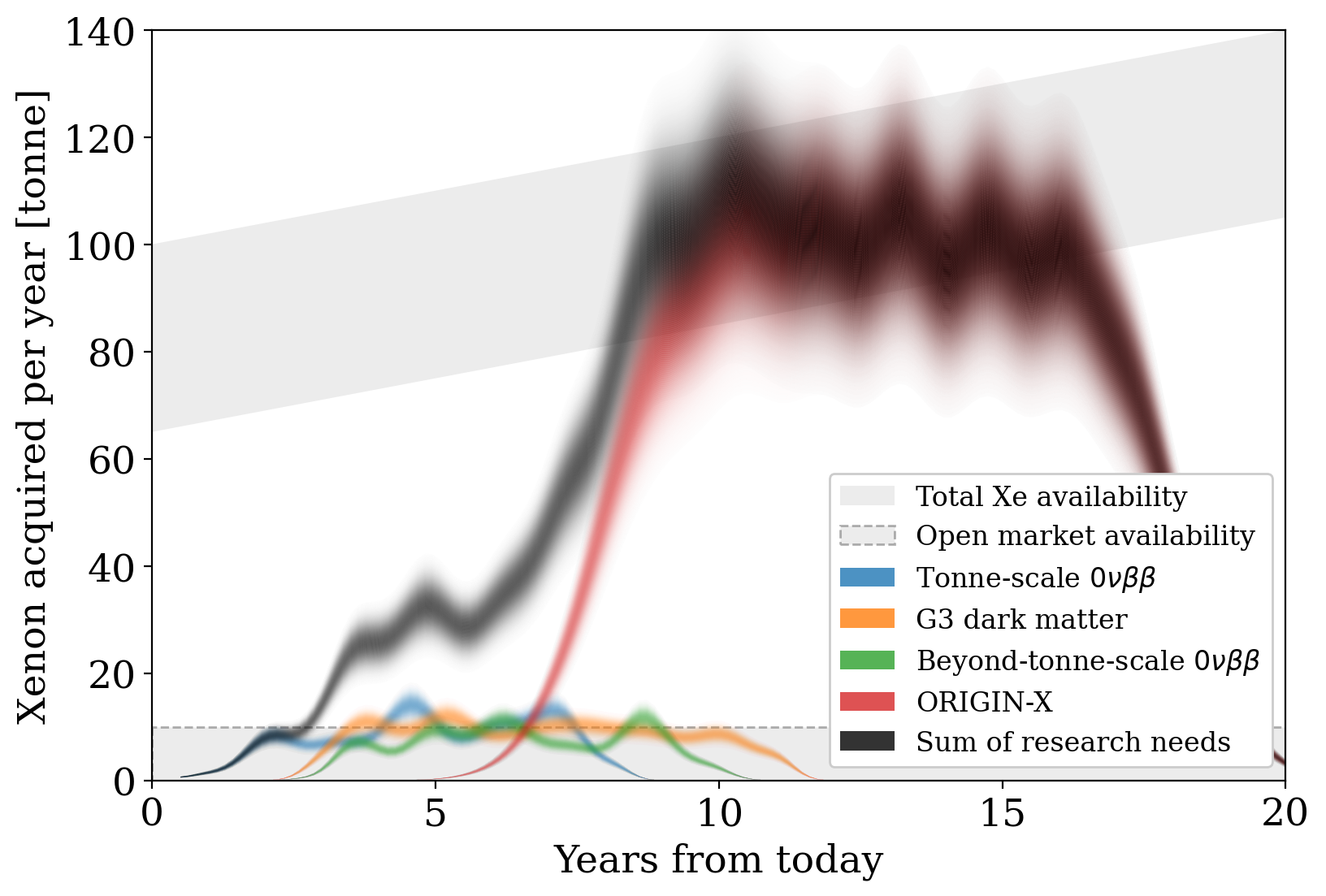}
    \caption{Rough projections of xenon requirements for various scientific programs in fundamental physics overlaid with the current availability on the open market and the entire world's supply. Here ``Beyond the tonne scale \0'' includes only programs using gaseous xenon TPCs. Additional proposals in which xenon is doped into other detector media are omitted; we discuss these possibilities in Section~\ref{subsec:synergies}. As the numbers used for this projection are estimates, the bands are shown without clear edges. }
    \label{fig:xenon_needs}
\end{figure}

Meanwhile, the need for xenon in fundamental physics research is growing, with multiple experiments on the horizon that will require tens of tonnes in the coming decades. We have summarized these needs in Figure~\ref{fig:xenon_needs}. Ongoing xenon programs to search for $0\nu\beta\beta$ decay, as a part of the ``tonne-scale'' program and beyond~\cite{AXEL,KZ,nEXO,NEXT,PANDAXIII}, will require several tonnes of enriched $^{136}$Xe, for which approximately $\sim$10$\times$ the amount in $^\mathrm{nat}$Xe must be processed. In parallel, the global effort to build a G3 dark matter experiment will require between 40-100 tonnes of $^\mathrm{nat}$Xe~\cite{Kamaha, Shutt}. The acquisition of xenon for these programs will occur over the next decade and is already at a scale that requires dedicated agreements with vendors beyond the availability on open markets. Further in the future, there are several proposals for experiments that will require $\mathcal{O}$(1000) tonnes of xenon to search for \0 -- we illustrate the ORIGIN-X concept in Fig.~\ref{fig:xenon_needs}. This level of xenon acquisition is beyond the capacity of the existing markets and therefore is likely impossible without substantial investment and/or additional xenon procurement pathways.

\subsection{Towards new xenon acquisition technologies}

To address these challenges, research and development efforts have begun to develop new technologies for xenon extraction from the air, known as Direct Air Capture (DAC). There are two key areas in which promising progress has been made, and in which further R\&D is necessary.

First is the development and characterization of new materials for high-capacity and high-selectivity xenon adsorption~\cite{Catarineu}. Though it is a noble gas and does not react chemically, xenon can be adsorbed due to its intrinsic atomic polarizability. Metal organic frameworks (MOFs) are a promising class of crystalline materials that can be engineered for the selectivity of specific gas species. MOFs are readily synthesized by combining a metal precursor with organic linkers; by careful selection of the organic linker the material's pore sizes can be tuned to achieve desired properties. Over the past few years, comprehensive studies of known MOFs have identified materials that improve upon commercial adsorbents for xenon capacity and selectivity. Initial characterization measurements of one of these materials, SBMOF-1, are promising~\cite{osti_1889654}. Further R\&D is needed to study long-term stability and performance when extracting xenon from the atmosphere under different conditions.

Second is the development of structured adsorbent beds and optimized processes to lower the cost of the adsorption process~\cite{Sangiorgio}. Parallel plane or 3D channel configurations can dramatically reduce the impedance of a bed compared to a more common packed-pellet configuration, reducing the energy required to push air through the apparatus. Furthermore, thermal swing cycles, as opposed to more common pressure swing cycles, may offer a significant cost reduction. Coupling these efforts into industrial-scale processes such as DAC for carbon sequestration offers an additional opportunity to further reduce the cost. An R\&D program is underway to study these using a prototype bed and detailed multiphysics simulations. While the R\&D program is still in its infancy, there appear to be promising paths toward the cost-effective production of xenon at the scales required for kilotonne-scale experiments.

\section{Detector program towards the kilotonne scale}

Prospects for the construction of a kilotonne-scale xenon detector (gas or liquid) come with new challenges and opportunities for advanced detector design. Readout planes grow with the surface area of the detector while the increase in volume contributes to increased attenuation/spreading of signals. Mechanical constraints significantly increase in difficulty due to radiopurity, size, and containment. The vast detector size also proves a challenge to efficiently circulate and purify the xenon while also being able to calibrate. 

The ideal background-free $0\nu\beta\beta$ decay experiment would incorporate barium tagging by unambiguously identifying a single barium ion in a kilotonne of material in coincidence with two electrons with a summed energy of 2.458 MeV. With the implementation of such a system, considerable effort may be required to incorporate it into the detector design. In addition, as discussed above, there may be a scenario where such a detector is designed such that the physics reach is extended to be sensitive to dark matter recoils. In any case, the challenge of the construction of a kilotonne-scale xenon detector is significant. 

\subsection{Detector Design Considerations}
Depending on the phase (i.e. liquid or gas), the challenges of building such a detector vary in its requirements. A gaseous detector, typically operated at high pressure, requires the containment of a large volume with significant stored energy. An exciting idea to solve this problem includes lined rock caverns~\cite{Monreal}. Such caverns have been developed for high-pressure gas storage suitable for 150-300 bar and entail a thin steel vessel lining the cavern walls with a layer of reinforced concrete between them. Under pressure, the load is transferred to the rock and therefore removes the need to build a separate containment vessel that requires significant material and engineering. The cost is relatively inexpensive and adds roughly \$100 per m$^3$ on top of the excavation of the cavern (existing caverns could also be used).

Lower-pressure gas is also an attractive option for improvement in the tracking due to the longer tracks. At the tonne scale and beyond, this option becomes viable due to the containment fraction of 1 MeV electrons although diffusion will need to be properly addressed due to the larger drift distance, most likely through the introduction of gas additives~\cite{GonzalezDiaz}. Mixtures of xenon with poly-atomic molecules such as CF$_4$ may be sufficient by rapidly cooling the ionization electrons through collisions. Benefits based on running with either a charge or light-based readout can be tuned with different concentrations. Additional studies into the compatibility of such molecules in sizeable concentrations with gas purifiers or effective mixing in larger volumes will be required.

For liquid-phase detectors, the primary advantage is the exclusion of radiogenic backgrounds via the strong self-shielding of liquid xenon, which improves exponentially with increasing detector size. At kilotonne scales, radiogenic backgrounds are so strongly attenuated that, at current material background levels, they are likely to be subdominant to uniformly-distributed backgrounds from e.g. solar neutrinos and $2\nu\beta\beta$. This means that achieving excellent energy resolution is crucial, requiring high light and charge collection efficiencies. In the specific case of solar neutrino backgrounds, it may be possible to utilize Cherenkov light signals to reduce their impact~\cite{Jamil}. In terms of detector design and construction, many challenges of kilotonne-scale noble-element time projection chambers are being addressed by ongoing programs (e.g. G3 dark matter and DUNE) with differences mostly amounting to radiogenics, xenon recovery, and purification. Scalable techniques to purify the xenon directly in the liquid phase have been developed for ongoing programs~\cite{refId0}. Removal of $^{222}$Rn, which emanates from surfaces in contact with the liquid, remains a topic of ongoing R\&D, with efforts including
direct $^{222}$Rn removal via distillation, higher-sensitivity screening of materials, mitigating entry into the active region (coatings, hermetic sealing, solid xenon), and tagging progeny decays~\cite{Shutt, Weinheimer}. The challenge of xenon storage and recovery at massive scales can be addressed with technologies like the ReStoX system, developed as a part of the XENON dark matter program~\cite{Masbou}.

Finally, solid xenon TPCs -- while still only demonstrated in laboratory-scale experiments -- are an exciting new development that may offer a powerful low-background detector technology for future experiments~\cite{Xia}.

\subsection{Instrumentation R\&D}
Current programs are developing new sensors for tonne-scale experiments, which could also apply to experiments at the kilotonne scale. Many avenues are being explored to maintain the stringent requirements for triggering, spatial resolution, energy resolution, and several mechanical/radiopurity constraints. While gas detectors using photomultiplier tubes (PMTs) have demonstrated excellent energy resolution ($<$1~\% full width at half maximum)~\cite{renner2019}, these devices are at present too radioactive for a beyond-the-next-generation \0 search~\cite{Mart_n_Albo_2016}. 

Readout methods such as dense tracking planes are being explored using standard SiPMs potentially combined with nanostructures such as metalenses to improve light collection efficiencies~\cite{Guenette} or digital SiPMs (consisting of Single-photon Avalanche Diodes (SPAD))~\cite{Keller, Charlebois}. The digital SiPMs integrate a single photon detector and its readout electronics on one silicon chip. This creates a system with purely digital output signals that reduce the power consumed and radioactivity that would normally accompany large arrays of SiPMs.

Alternative readout methods that may be easy to scale include fast-optical cameras with a single photon-sensitive image intensifier~\cite{Rogers}, electroluminescent light collection cells~\cite{Akiyama}, or field-assisted transparent gaseous electron multipliers~\cite{Leardini}. 

\subsection{Barium Tagging}
In the \0 decay of $^{136}$Xe, two electrons with a summed energy at the Q-value and a charged barium daughter ion are produced. Present searches use a combination of energy and topology of the electrons to remove backgrounds, however, these techniques are prone to contamination of radiogenic and cosmic backgrounds. \0 experiments require less than 1 count per tonne-year, so even small contaminations are detrimental to the experimental sensitivity. The most promising solution to completely remove these backgrounds includes tagging the barium in coincidence with the electron signal.

Prospects for barium tagging are being explored in liquid and gaseous phase TPC detectors, led by the nEXO and NEXT collaborations respectively. The roadmap contains an intense R\&D program to realize the technological components, followed by demonstrator phases and implementation in tonne scale programs such as nEXO and NEXT-HD. 

From the creation of the barium ion, identification of its location and trigger is obtained through the measurement of the ionization electrons energy signal, which has a faster drift time. The barium ion can then be brought to the location of the sensing equipment. While detection schemes are not completely determined due to ongoing R\&D along many avenues, we summarize here the latest progress of three programs. 

The current scheme of the Canadian Barium Tagging Concept (CBTC)~\cite{Brunner} involves extracting the barium with capillary transferring from liquid to gas phase, moving the ion to a radio frequency (RF) funnel which separates the ions from neutral particles into a Linear Paul Trap (LPT). The barium is then detected with laser fluorescence spectroscopy scanning over the ions within the LPT. A final step includes a multiple reflection time-of-flight (MRTOF) measurement that determines the mass of the barium for systematic control. The program is currently in the development of a LXe barium ion source for studying the extraction efficiency. Progress has been made towards a test stand for studying the ion extraction, and constructing an LPT and MRTOF with all pieces under the commissioning and testing stages. In addition, an RF funnel extraction system has been developed. 

The solid xenon approach~\cite{Fairbank} employs the use of a cryogenic probe to freeze the liquid xenon in the expected region of the barium ion to extract and detect its presence with laser-induced fluorescence. Excellent progress has been made in understanding the fluorescence of the barium, its bleaching properties in the solid xenon matrix, and the backgrounds in the imaging process. In addition, an apparatus for capturing the barium in solid xenon has been developed.

The method of single-molecule fluorescence imaging (SMFI)~\cite{Jones} is being developed for a gas-phase detector and is based on the concept of turning a non-fluorescent molecule into a fluorescent one upon the chelation of a barium ion (or similar variants). The method employs the use of RF carpets~\cite{Brodeur} to sweep the barium ions directly to a sensor that contains the sensing molecules. This program has made progress in constructing molecules that work in the dry phase, as well as additional types that have different fluorescent properties such as a color shift. A high-pressure microscopy system with automatic focusing has also been demonstrated and fine-pitched four-phased RF carpets are being tested. A new concept is also being explored which integrates the RF electronics and barium sensors into a single integrated chip that can be manufactured at scale and placed across the detector wall. 

\subsection{Synergies with other efforts in fundamental physics}
\label{subsec:synergies}

In addition to the ORIGIN-X concept, there are proposals from the long-baseline neutrino community for beyond-the-next-generation experiments using large quantities of xenon. The liquid argon TPCs used by DUNE could be doped with xenon at the few-percent level, which would provide improved scintillation response and enhance the standard physics program; recent work has shown that, with careful low-background design, such a detector could also provide competitive sensitivity to \0~\cite{Zennamo,Psihas}. Similarly, large-scale liquid scintillator detectors such as JUNO or the proposed THEIA experiment~\cite{Kaptanoglu}, with \%-level doping of $^{136}$Xe, could provide excellent sensitivity to \0. Both cases would require $\mathcal{O}(100)$ tonnes of xenon, and could be envisioned as intermediate steps in a kilotonne-scale program: as the community acquires more and more xenon, it could be added as a dopant to these detectors, continuing to push the science forward while also ``storing'' it until sufficient quantities are procured to build a dedicated xenon double beta decay experiment. These types of synergy would require sophisticated coordination between different experimental communities. Likewise, the G3 dark matter effort may provide a compelling path towards a $\sim$100 tonne \0 experiment. This program shares many of the low-background requirements with \0 searches~\cite{Shutt}, and specific requirements for dark matter detectors -- e.g., low thresholds (e.g. single electron sensitivity) and high-light collection efficiency -- do not contradict the requirements in a \0 detector, although the choice in instrumentation may require compromise. Assuming that design choices can be optimized for both science goals, one could imagine a G3 dark matter search which, after its primary WIMP search program and initial \0 results, is ``upgraded'' by filling with xenon enriched in $^{136}$Xe for a 100-tonne-scale \0 search 
without requiring any significant overhauls to the detector design. 

These ideas illustrate the potential for a community-wide, phased approach to continue the search for new physics as the stockpile of xenon is accumulated, and should continue to be explored.

\section{Summary and Outlook}

The search for neutrinoless double beta decay is one of the most promising avenues to search for new physics, with dramatic implications for fundamental physics and cosmology. The isotope $^{136}$Xe is particularly attractive, due to the possibility of using it in a wide variety of highly scalable, low-background detector technologies with the potential to achieve unprecedented exposures and sensitivities. Equally important is that many of the necessary technologies for realizing beyond-the-next-generation \0 searches are being developed in parallel with other programs in fundamental physics, and cross-cutting R\&D can be leveraged to maximize the scientific impact of future experiments. The \emph{Workshop on Xenon Detector Double Beta Decay Searches: Steps Towards the Kilotonne Scale} brought together experts across a wide range of experiments and programs, with the goal of fostering collaboration among scientists working towards the common goal of beyond-the-next-generation experiments searching for double beta decay and other rare phenomena. While many exciting challenges lie ahead, the community is developing the techniques and technologies to enable powerful exploration of new physics in the coming decades.



\bibliographystyle{apsrev}
\bibliography{biblio}

\end{document}